\documentclass[conference,draftcls,onecolumn]{IEEEtran}	% SITE

\usepackage[latin1]{inputenc}
\usepackage[T1]{fontenc}
\usepackage[english]{babel}
\usepackage{amssymb,amsmath,amsfonts}
\usepackage{times}
\usepackage{graphicx}
\usepackage{color}
\usepackage{verbatim}
\usepackage{cite}

\usepackage{subfig}	
\usepackage{pgf}		
\usepackage{tikz}
\usetikzlibrary{decorations}
\usetikzlibrary{patterns}
\usetikzlibrary{positioning}
\usetikzlibrary{backgrounds}

\usepackage{stmaryrd}

\usepackage{cancel}
\usepackage[normalem]{ulem}

\newtheorem{defi}	{Definition}
\newtheorem{theo}	{Theorem}
\newtheorem{prop}	{Proposition}

\newtheorem{remark}	{Remark}

\newcommand{\cA}{{\mathcal A}}
\newcommand{\cB}{{\mathcal B}}
\newcommand{\cE}{{\mathcal E}}
\newcommand{\cQ}{{\mathcal Q}}
\newcommand{\cR}{{\mathcal R}}

\newcommand{\cT}{{\mathcal T}}
\newcommand{\cU}{{\mathcal U}}
\newcommand{\cV}{{\mathcal V}}
\newcommand{\cX}{{\mathcal X}}
\newcommand{\cY}{{\mathcal Y}}
\newcommand{\cZ}{{\mathcal Z}}

\newcommand{\bN}{{\mathbb N}}
\newcommand{\bR}{{\mathbb R}}
\newcommand{\bE}{{\mathbb E}}

\newcommand{\pr}[1]{\operatorname{Pr}\left\{#1\right\}}

\newcommand{\mkv}{-\!\!\!\!\minuso\!\!\!\!-}

\newcommand{\lessnoisy}[1]{\succeq_{\scriptscriptstyle #1}}
\usepackage{hyperref}			% SITE
\IEEEoverridecommandlockouts	% pour faire \thanks

\begin{document}

\title{	Secure Lossy Source-Channel Wiretapping with Side Information at the Receiving Terminals} 

\author{
\IEEEauthorblockN{
	Joffrey Villard\IEEEauthorrefmark{1},
	Pablo Piantanida\IEEEauthorrefmark{1} and
	Shlomo Shamai (Shitz)\IEEEauthorrefmark{2}
}
\\
\IEEEauthorblockA{
\begin{tabular}{cc}
	\IEEEauthorrefmark{1}	Department of Telecommunications	&
	\IEEEauthorrefmark{2}	Department of Electrical Engineering	\\
	SUPELEC										&	Technion - Israel Institute of Technology\\	
	91192 Gif-sur-Yvette, France				&	Technion city, Haifa 32000, Israel\\
	Email: \{joffrey.villard, pablo.piantanida\}@supelec.fr &
	Email: sshlomo@ee.technion.ac.il	
\end{tabular}
}

\thanks{The work of J. Villard is supported by DGA (French Armament Procurement Agency). This research is partially supported by the FP7 Network of Excellence in Wireless COMmunications NEWCOM++.}
}

\date{May 2011} % SITE 

\maketitle
 
\renewcommand{\leftmark}{\MakeUppercase{to be presented at ISIT 2011}}		% SITE
\renewcommand{\rightmark}{} 												% SITE
 
%--------------------------------------------------------------------
\begin{abstract}
The problem of secure lossy source-channel wiretapping with arbitrarily correlated side informations at both receivers is investigated. This scenario consists of an encoder (referred to as Alice) that wishes to compress a source and send it through a noisy channel to a legitimate receiver (referred to as Bob). In this context, Alice must simultaneously satisfy the desired requirements on the distortion level at Bob, and the equivocation rate at the eavesdropper (referred to as Eve). 
This setting can be seen as a generalization of the conventional problems of secure source coding with side information at the decoders, and the wiretap channel. Inner and outer bounds on the rate-distortion-equivocation region for the case of arbitrary channels and side informations are derived. 
In some special cases of interest, it is shown that separation holds.
By means of an appropriate coding, the presence of any statistical difference among the side informations, the channel noises, and the distortion at Bob can be fully exploited in terms of secrecy. 
\end{abstract}

%==============================================================================
\section{Introduction}

Consider a system composed of three nodes (or sensors) where each one is measuring an analogue source (or random field) as a function of time. In order to make reliable decisions, one of these sensors (referred to as Bob) can be helped by another one (referred to as Alice), which will transmit some compressed version of its own measurement through a noisy wireless channel. The third sensor (referred to as Eve) can listen to the wireless medium, and capture some information during the communication. Considering that Eve is not to be trusted (she is an \emph{eavesdropper}), Alice wishes to leak the least possible amount of information about its source.

The above scenario involves most of the major information-theoretic issues on (secure) source and channel coding. In fact, the information-theoretic notion of secrecy was first introduced by Shannon in~\cite{shannon1949communication}, where security is measured through the equivocation rate, \emph{i.e.}, the remaining uncertainty about the message, at Eve. 
In terms of source coding, Slepian and Wolf~\cite{slepian1973noiseless}, and Wyner and Ziv~\cite{wyner1976rate} introduced the problem of source coding with side information at the decoder. The corresponding \emph{secure} scenarios \emph{i.e.}, involving an eavesdropper with its own side information, have been recently studied in~\cite{prabhakaran2007secure,gunduz2008secure,gunduz2008lossless,tandon2009securea,villard2010secure}. 
Secure source coding scenarios involving a secure rate-limited channel between Alice and Bob, which allows the use of secret keys, have also been studied in various works~\cite{yamamoto1994coding,yamamoto1997rate,liu2009securing,merhav2006shannon}. 
On the other hand, extensive research has been done during the recent years on secure communications over noisy channels. The wiretap channel was introduced by Wyner~\cite{wyner1975wire}, who showed that it is possible to send information with perfect secrecy as long as the channel of Bob is less noisy than the channel of Eve. Csisz\`ar and K\''orner~\cite{csiszar1978broadcast} extend this result to the setting of general broadcast channels with arbitrary equivocation rate (allowing also a common message to both receivers). Several extensions of the wiretap channel have since been done (cf. \cite{yamamoto1997rate,merhav2008shannon,it2008special,liang2009information} and references therein).
Whereas, secure lossy source-channel coding problems have received fewer attention. In a recent work~\cite{merhav2008shannon}, Merhav considered such a setting by assuming that Eve has a degraded channel with degraded side information with respect to Bob, and that a secret key can be shared between Alice and Bob.

In this paper, we investigate the general problem of secure lossy source-channel wiretapping, with arbitrarily correlated side informations as depicted in Fig.~\ref{fig:schema}. 
The main goal is to understand how Alice can take advantage of the presence of statistical differences among the side informations and the channel noises to reveal the minimum amount of information to Eve, and satisfy the required distortion level at Bob. 
It should be emphasized that the central difficulty of this problem lies in the evaluation of the equivocation at Eve.
We derive single-letter characterizations of inner and outer bounds on the general rate-distortion-equivocation region (in Section~\ref{sec:main}). 
Section~\ref{sec:special} provides special cases for which separation holds. The sketches of the proofs are relegated to Sections~\ref{sec:inner_bound} and~\ref{sec:outer_bound}. Finally, Section~\ref{sec:example} presents discussions and an application example to binary sources.

%------------------------------------------------------------------------------
\subsection*{Notations}
For any sequence~$(x_i)_{i\in\bN^*}$, notation $x_k^n$
stands for the collection $(x_k,x_{k+1},\dots, x_n)$.
$x_1^n$ is simply denoted by $x^n$.
Entropy is denoted by $H(\cdot)$, and mutual information by $I(\cdot;\cdot)$.
Let $X$, $Y$ and $Z$ be three random variables on some alphabets with probability distribution~$p$.
If $p(x|y,z)=p(x|y)$ for each $x,y,z$, then they form a Markov chain, which is denoted by $X\mkv Y\mkv Z$.
The set of nonnegative real numbers is denoted by $\bR_+$.
For each $x\in\bR$, notation $[x]_+$ stands for $\max(0;x)$.

%==============================================================================
\section{Problem Definition and Main Results}
\label{sec:main}

%::::::::::::::::::::::::::::::::::::::
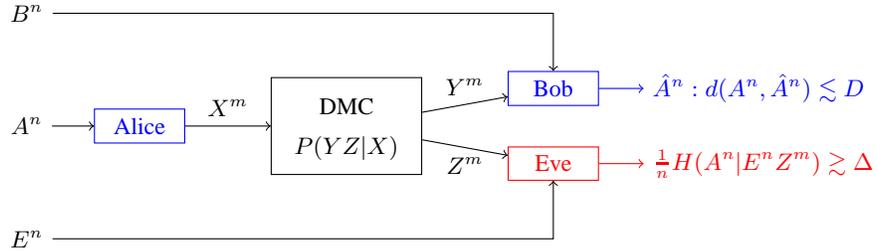
\begin{figure}
\centering
\begin{tikzpicture}
	\tikzstyle{user}=[rectangle,draw,minimum width = 1.2cm];

	\node	(A) 		at (0,0) 					 					{\small $A^n$};
	\node	(alice)		at (1.5,0)	[user,blue] 						{\small Alice};
	\node	(canal)		at (4.25,0)	[rectangle,draw,minimum width=2cm] 	{\small \begin{tabular}{c}DMC\\ $\!\!\!\!\! P(YZ|X)\!\!\!\!\!$\end{tabular}};
	\node	(B) 		at (0,1.5) 					 					{\small $B^n$};
	\node	(bob) 		at (7,.5)	[user,blue] 						{\small Bob};
	\node	(hatA) 		at (8.2,.5)	[right,blue]	 					{\small $\hat A^n: d(A^n,\hat A^n)\lesssim D$};
	\node	(eve) 		at (7,-.5)[user,red]							{\small Eve};
	\node	(E) 		at (0,-1.5)					 					{\small $E^n$};
	\node	(D) 		at (8.2,-.5)[right,red]							{\small $\frac1n H(A^n|E^n Z^m)\gtrsim\Delta$};
		
	\draw[->]	(A)			to (alice);
	\draw[->]	(B)			to (7,1.5)	to  (bob);
	\draw[->]	(E)			to (7,-1.5)	to (eve);
	
	\draw[->]	(alice)		to node[anchor=south]{\small $X^m$} (canal);
	\draw[->]	(canal)		to node[anchor=south]{\small $Y^m$} (bob);
	\draw[->]	(canal)		to node[anchor=north]{\small $Z^m$} (eve);
	
	\draw[->,blue]	(bob)	to (hatA);
	\draw[->,red]	(eve)	to (D);
\end{tikzpicture}
\caption{Secure lossy source-channel wiretapping in the presence of side information at the receivers.}
\label{fig:schema}\vspace{-4mm}
\end{figure}
%::::::::::::::::::::::::::::::::::::::

%------------------------------------------------------------------------------
\subsection{Problem Definition}

In this section, we give a more rigorous formulation of the context depicted in Fig.~\ref{fig:schema}.
Let $\cA$, $\cB$, $\cE$, $\cX$, $\cY$, and $\cZ$ be six finite sets. 
Alice, Bob, and Eve observe the sequences of random variables 
$(A_i)_{i\in\bN^*}$, $(B_i)_{i\in\bN^*}$, and $(E_i)_{i\in\bN^*}$,
respectively, which take values on $\cA$, $\cB$, and $\cE$, resp.
For each $i\in\bN^*$, the random variables $A_i$, $B_i$, and $E_i$
are distributed according to the joint distribution $p(a,b,e)$ on
$\cA\times\cB\times\cE$.
Moreover, they are independent across time $i$.
Alice can also communicate with Bob and Eve through a discrete memoryless channel with input $X$ on $\cX$, and outputs $Y$, $Z$ on $\cY$, $\cZ$, respectively. This channel is defined by its transition probability $P(YZ|X)$.

Let $d : \cA\times\cA \to [0\,;d_{\max}]$ be a finite distortion measure
\emph{i.e.}, such that $0\leq d_{\max} < \infty$.
We also denote by $d$ the component-wise mean distortion on $\cA^n\times\cA^n$
\emph{i.e.}, for each $a^n,b^n\in\cA^n$, $d(a^n,b^n) = \frac1n\,\sum_{i=1}^n d(a_i,b_i)$.

\begin{defi}
An $(n,m)$-code for source-channel coding in this setup is defined by
\begin{itemize}
\item A (stochastic) encoding function at Alice $F : \cA^n \to \cX^m$, defined by some transition probability $P_{X^m|A^n}(\cdot|\cdot)$,
\item A decoding function at Bob $g : \cY^m\times\cB^n \to \cA^n$.
\end{itemize}
\end{defi}

The rate of such a code is defined as quantity $m/n$ (\emph{channel uses per source symbol}).

\begin{defi}
A tuple $(k,D,\Delta)\in\bR_+^3$ is said to be \emph{achievable} if,
for any $\varepsilon>0$, there exists an $(n,m)$-code $(F,g)$ s.t.:
\begin{IEEEeqnarray*}{rCl}
\frac mn 							&\leq& k+\varepsilon \ ,\\
\bE\big[ d(A^n,g(Y^m,B^n)) \big]	&\leq& D+\varepsilon \ ,\\
\dfrac1n\,H(A^n|E^n Z^m) 			&\geq& \Delta-\varepsilon \ ,
\end{IEEEeqnarray*}
when the input of the channel $X^m$ is the output of the encoder $F(A^n)$.
The set of all achievable tuples is denoted by $\cR^*$
and is referred to as the \emph{rate-distortion-equivocation region}.
\end{defi}

%------------------------------------------------------------------------------
\subsection{Main Results}

% - - - - - - - - - - - - - - - - - - - - - - - - - - - - - - - - - - - - - - -
The following theorem gives an inner bound on $\cR^*$ \emph{i.e.}, it defines region $\cR_\text{in}\subset\cR^*$. The proof is outlined in Section~\ref{sec:inner_bound}.

\begin{theo}[Inner Bound]
\label{th:inner_bound}
The set of all tuples $(k,D,\Delta)$ in $\bR_+^3$ such that there exist random variables $U$, $V$, $Q$, $T$ on some finite sets $\cU$, $\cV$, $\cQ$, $\cT$, respectively, with joint distribution $p(uvqtabexyz) = p(u|v)p(v|a)p(abe)p(q|t)\linebreak p(t|x)p(xyz)$, and a function $\hat A : \cV\times\cB \to \cA$, verifying the following inequalities, is achievable:
\begin{IEEEeqnarray*}{rCl}
I(U;A|B) &\leq& k I(Q;Y) 						\ ,\\
I(V;A|B) &\leq& k I(T;Y) 						\ ,\\
D 		 &\geq& \bE\big[d(A,\hat A(V,B))\big]	\ ,\\
\Delta	 &\leq&  H(A|UE) - \Big[ I(V;A|UB) - k \Big( I(T;Y|Q) - I(T;Z|Q) \Big) \Big]_+ .
\end{IEEEeqnarray*}
\end{theo}

The first two inequalities in Theorem~\ref{th:inner_bound} correspond to sufficient conditions for the transmission of two source layers $U$, $V$ in channel variables $Q$, $T$, resp. 
The first layer $(U\mapsto Q)$ can be seen as a \emph{common} message which is considered to be known at Eve, as shown by the term $H(A|UE)$ in the equivocation.
The second layer $(V\mapsto T)$ forms a \emph{private} message which is (partially) protected by adding an independent random noise~\cite{csiszar1978broadcast,liang2009information}. 
The term inside the brackets in the fourth inequality corresponds to the information that Eve can still obtain on this protected layer.

% - - - - - - - - - - - - - - - - - - - - - - - - - - - - - - - - - - - - - - -
The following theorem gives an outer bound on $\cR^*$ \emph{i.e.}, it defines region $\cR_\text{out}\supset\cR^*$. The proof is outlined in Section~\ref{sec:outer_bound}.

\begin{theo}[Outer Bound]
\label{th:outer_bound}
For each achievable tuple $(k,D,\Delta)$, there exist random variables $U$, $V$, $Q$, $T$ on some finite sets $\cU$, $\cV$, $\cQ$, $\cT$, respectively, and a function $\hat A : \cV\times\cB \to \cA$, such that $p(uvqtabexyz) = p(uv|a)p(abe)\linebreak p(q|t)p(t|x)p(xyz)$, and
\begin{IEEEeqnarray*}{rCl}
I(V;A|B) &\leq& k I(T;Y) 									\ ,\\
D 		 &\geq& \bE\big[d(A,\hat A(V,B))\big] 			\ ,\\
\Delta	 &\leq&  H(A|UE) - \Big[ I(V;A|B) - I(U;A|B) - k \Big( I(T;Y|Q) - I(T;Z|Q) \Big) \Big]_+ .
\end{IEEEeqnarray*}
\end{theo}

% - - - - - - - - - - - - - - - - - - - - - - - - - - - - - - - - - - - - - - -
Notice that the inner and outer bounds do not meet in general. In Section~\ref{sec:special}, we provide several cases where $\cR_\text{in}$ is optimal. In fact, there are two main differences between $\cR_\text{in}$ and $\cR_\text{out}$:
\begin{itemize}
\item The first inequality of Theorem~\ref{th:inner_bound}, which is needed in our scheme to characterize the equivocation at Eve, may not be optimal for the general case,
\item The Markov chain $U\mkv V\mkv A\mkv(B,E)$ is assumed in Theorem~\ref{th:inner_bound} while only $(U,V)\mkv A\mkv(B,E)$ is proved for arbitrary codes in Theorem~\ref{th:outer_bound}.
\end{itemize}

\begin{figure}
\centering
%::::::::::::::::::::::::::::::::::::::
\begin{minipage}[t]{8cm}
\centering
\begin{tikzpicture}[scale=.8]
	\node	(A) 		at (0,0) 					 	{\small $A^n$};
	\node	(aliceS)	at (2,0)	[rectangle,draw] 	{\small \begin{tabular}{c}Source\\ encoder\end{tabular}};		
	\node	(aliceC)	at (5.8,0)	[rectangle,draw] 	{\small \begin{tabular}{c}Channel\\ encoder\end{tabular}};
	\node	(canal)		at (8,0)	 					{\small $X^m$};
		
	\draw[->]	(A)			to (aliceS);
	\draw[->]	(aliceS)	to node[anchor=south]{\small $r$} (aliceC);
	\draw[->]	(aliceC)	to  (canal);
\end{tikzpicture}
\caption{Traditional (``informational'') separation.}
\label{fig:separation}
\end{minipage}
%::::::::::::::::::::::::::::::::::::::
\begin{minipage}[t]{8cm}
\centering
\begin{tikzpicture}[scale=.8]
	\node	(A) 		at (0,0) 					 	{\small $A^n$};
	\node	(aliceS)	at (2,0)	[rectangle,draw] 	{\small \begin{tabular}{c}Source\\ encoder\end{tabular}};		
	
	\node	(aliceS1)	at (2.9,.3)						{};			
	\node	(aliceS2)	at (2.9,-.3)					{};
			
	\node	(aliceC1)	at (4.9,.3)						{};			
	\node	(aliceC2)	at (4.9,-.3)					{};
	
	\node	(aliceC)	at (5.8,0)	[rectangle,draw] 	{\small \begin{tabular}{c}Channel\\ encoder\end{tabular}};
	\node	(canal)		at (8,0)	 					{\small $X^m$};
		
	\draw[->]	(A)			to (aliceS);
	\draw[->]	(aliceS1)	to node[anchor=south]{\small $r_c$} (aliceC1);
	\draw[->]	(aliceS2)	to node[anchor=north]{\small $r_p$} (aliceC2);
	\draw[->]	(aliceC)	to  (canal);
\end{tikzpicture}
\caption{Proposed system (``operational'' separation).}
\label{fig:op_separation}\vspace{-4mm}
\end{minipage}
%::::::::::::::::::::::::::::::::::::::
\end{figure}

%------------------------------------------------------------------------------
\subsection{Coding Scheme Based on ``Operational'' Separation}

In traditional \emph{separated} schemes, two \emph{stand-alone} components successively perform source and channel coding, as depicted in Fig.~\ref{fig:separation}. However the proposed scheme (which achieves region~$\cR_\text{in}$) does not satisfy this separation principle: The source encoder outputs two layers (as in~\cite{villard2010secure}) which are further encoded by using the channel code for a \emph{broadcast channel with a confidential message}~\cite{csiszar1978broadcast}. This results in two independent (but not stand-alone) source and channel components leading to statistically independent source and channel variables (as in~\cite{tuncel2006slepian-wolf} for Slepian-Wolf coding over broadcast channels) \emph{i.e.}, ``operational'' separation holds (see Fig.~\ref{fig:op_separation}).
As a matter of fact, the first inequality of Theorem~\ref{th:inner_bound} \emph{i.e.}, $I(U;A|B)\leq k I(Q;Y)$, prevents from separately choosing variables $U$ and $Q$ which would maximize the equivocation rate at Eve.

%==============================================================================
\section{Special Cases of Interest}
\label{sec:special}

In this section, we characterize the optimality of the inner bound $\cR_\text{in}$ for some special cases.

%------------------------------------------------------------------------------
\subsection{Bob Has Less Noisy Side Information}

\begin{defi}
Random variable $B$ is \emph{less noisy} than $E$ w.r.t. $A$, if $I(U;B) \geq I(U;E)$ for each r.v. $U$ s.t. $U\mkv A\mkv (B,E)$ form a Markov chain.
This relation is denoted by $B \lessnoisy{A} E$.
\end{defi}

% - - - - - - - - - - - - - - - - - - - - - - - - - - - - - - - - - - - - - - -
\begin{prop}
\label{prop:B_less_noisy}
If $B \lessnoisy{A} E$, then region $\cR^*$ reduces to the set of all tuples $(k,D,\Delta)\in\bR_+^3$ such that there exist random variables $V$, $Q$, $T$ on some finite sets $\cV$, $\cQ$, $\cT$, respectively, with joint distribution $p(vqtabexyz) = p(v|a)p(abe)p(q|t)p(t|x)p(xyz)$, and a function $\hat A : \cV\times\cB \to \cA$, verifying the following inequalities:
\begin{IEEEeqnarray*}{rCl}
I(V;A|B) &\leq& k I(T;Y) 							\ ,\\
D 		 &\geq& \bE\big[d(A,\hat A(V,B))\big] 	\ ,\\
\Delta	 &\leq&  H(A|E) - \Big[ I(V;A|B) - k \Big( I(T;Y|Q) - I(T;Z|Q) \Big) \Big]_+ .
\end{IEEEeqnarray*}
\end{prop}

\begin{remark}
In this case, the optimal coding reduces to a \emph{Wyner-Ziv} source encoder~\cite{wyner1976rate} followed by a classical \emph{wiretap} channel encoder~\cite{csiszar1978broadcast,liang2009information}, and hence the conventional separation principle holds (Fig.~\ref{fig:separation}).
\end{remark}

% - - - - - - - - - - - - - - - - - - - - - - - - - - - - - - - - - - - - - - -
\begin{IEEEproof}
The above region is achievable by setting variable $U$ to a constant value in Theorem~\ref{th:inner_bound}.
On the other hand, the third inequality of Theorem~\ref{th:outer_bound} writes:
\begin{IEEEeqnarray*}{rCl}
\Delta	&\leq&  H(A|UE) 						\\
\Delta	&\leq&  H(A|VB) + I(A;B|U) - I(A;E|U) - k \Big( I(T;Y|Q) - I(T;Z|Q) \Big)	\ .
\end{IEEEeqnarray*}
Since $B \lessnoisy{A} E$, and $U\mkv A\mkv(B,E)$ form a Markov chain, $I(A;B|U) - I(A;E|U) \leq I(A;B) - I(A;E)$. Moreover $H(A|UE)\leq H(A|E)$. In this case, the outer bound $\cR_\text{out}$ is thus included in (and consequently equal to) $\cR_\text{in}$.
\end{IEEEproof}

% - - - - - - - - - - - - - - - - - - - - - - - - - - - - - - - - - - - - - - -
If the informations at Eve (both side information, and channel output) are degraded versions of Bob's ones \emph{i.e.}, if both Markov chains $A\mkv B\mkv E$, and $X\mkv Y\mkv Z$ hold, then Proposition~\ref{prop:B_less_noisy} reduces to the results in~\cite{merhav2008shannon}. In this case, variable $Q$ is set to a constant value, and $T=X$.

%------------------------------------------------------------------------------
\subsection{Eve Has Less Noisy Channel Output}

% - - - - - - - - - - - - - - - - - - - - - - - - - - - - - - - - - - - - - - -
\begin{prop}
\label{prop:Z_less_noisy}
If $Z \lessnoisy{X} Y$, then region $\cR^*$ reduces to the set of all tuples $(k,D,\Delta)\in\bR_+^3$ such that there exist random variables $U$, $V$ on some finite sets $\cU$, $\cV$, respectively, with joint distribution $p(uvabexyz) = p(u|v)\linebreak p(v|a)p(abe)p(xyz)$, and a function $\hat A : \cV\times\cB \to \cA$, verifying the following inequalities:
\begin{IEEEeqnarray*}{rCl}
I(V;A|B)&\leq& k I(X;Y) 						\ ,\\
D 		&\geq& \bE\big[d(A,\hat A(V,B))\big] 	\ ,\\
\Delta	&\leq&  H(A|VB) + I(A;B|U) - I(A;E|U) 	\ .
\end{IEEEeqnarray*}
\end{prop}

\begin{remark}
In this case, the optimal scheme reduces to a \emph{secure} source encoder~\cite{villard2010secure} followed by a conventional channel encoder, and hence separation principle holds (Fig.~\ref{fig:separation}).
\end{remark}

% - - - - - - - - - - - - - - - - - - - - - - - - - - - - - - - - - - - - - - -
\begin{IEEEproof}
The above region is achievable by setting $Q=T=X$ in Theorem~\ref{th:inner_bound}. 
However, a new proof is needed to obtain the converse part of Proposition~\ref{prop:Z_less_noisy}. Here, auxiliary variables are defined as follows, for each $i\in\{1,\dots,n\}$, and each $j\in\{1,\dots,m\}$:
\begin{IEEEeqnarray*}{rCl}
U_i &=& (\phantom{A^{i-1},B^{i-1}, } B_{i+1}^n,E^{i-1}, Y^m)	\ ,\\
V_i &=& (A^{i-1},B^{i-1},			 B_{i+1}^n,E^{i-1}, Y^m)	\ ,\\[.1cm]
Q_j &=& (\phantom{A^n, }	E^n, Y^{j-1}, Z_{j+1}^m) 			\ ,\\
T_j &=& (A^n,				E^n, Y^{j-1}, Z_{j+1}^m) 			\ .
\end{IEEEeqnarray*}
Now, both $U_i\mkv V_i\mkv A_i\mkv (B_i,E_i)$, and $Q_j\mkv T_j\mkv X_j\mkv (Y_j,Z_j)$ form Markov chains. 
Following the arguments given at Section~\ref{sec:outer_bound}, we can define new auxiliary variables verifying the above Markov chains and the following inequalities:
\begin{IEEEeqnarray*}{rCl}
I(V;A|B) &\leq& k I(T;Y) 							\ ,\\
D 		 &\geq& \bE\big[d(A,\hat A(V,B))\big] 		\ ,\\
\Delta	 &\leq&  H(A|UE) - I(V;A|UB) + k \Big( I(T;Y|Q) - I(T;Z|Q) \Big) .
\end{IEEEeqnarray*}
Since $Z\lessnoisy{X}Y$, and $Q\mkv T\mkv X\mkv(Y,Z)$ form a Markov chain, $I(T;Y|Q) - I(T;Z|Q)\leq0$. Noting that $I(T;Y)\leq I(X;Y)$, this concludes the proof.
\end{IEEEproof}

% - - - - - - - - - - - - - - - - - - - - - - - - - - - - - - - - - - - - - - -
Defining the transmitted rate as $R=kI(X;Y)$, Proposition~\ref{prop:Z_less_noisy} provides the \emph{rate-distortion-equivocation} region in the secure \emph{source} coding setup~\cite[Theorem~1]{villard2010secure}.

%==============================================================================
\section{Sketch of Proof of Theorem~\ref{th:inner_bound} (Inner Bound)}
\label{sec:inner_bound}

The proof is based on the use of a secure source coding scheme~\cite{villard2010secure}, and a channel coding scheme for wiretap channel~\cite{csiszar1978broadcast,liang2009information}. Full details are omitted due to the lack of space and will be provided in an extended version of this paper.

% - - - - - - - - - - - - - - - - - - - - - - - - - - - - - - - - - - - - - - -
\subsubsection*{Source Encoder}
The source encoder is formed of two layers corresponding to variables $U$, $V$, with respective rates $R_1$, $R_2$.
Random binning \emph{a la} Wyner-Ziv~\cite{wyner1976rate} is performed prior to transmission. The next constraints ensure that Bob can decode $(U,V)$ from bin indices $(r_1,r_2)$ with an arbitrarily small error probability:
\begin{IEEEeqnarray*}{rCl}
R_1 &>& I(U;A|B)	\ ,\\
R_2 &>& I(V;A|UB)	\ .
\end{IEEEeqnarray*}

% - - - - - - - - - - - - - - - - - - - - - - - - - - - - - - - - - - - - - - -
\subsubsection*{Bits Recombination}
Bin indices $(r_1,r_2)$ are mapped to indices $r_c$ and $r_p$, with respective rates $R_c$, $R_p$, through a one-to-one mapping, such that $r_1 = M'(r_c)$ for some mapping $M'$. This requires the following constraints:
\begin{IEEEeqnarray*}{rCl}
R_1 + R_2 	&=& 	R_c + R_p	\ ,\\
R_1 		&\leq& 	R_c 		\ .
\end{IEEEeqnarray*}

% - - - - - - - - - - - - - - - - - - - - - - - - - - - - - - - - - - - - - - -
\subsubsection*{Channel Encoder}
The channel encoder is composed of two layers corresponding to variables $Q$, $X$, transmitting messages $r_c$, $r_p$, respectively. Following~\cite{csiszar1978broadcast,liang2009information}, an independent random noise $r_f$, with rate $R_f$ s.t. $R_f < k I(X;Z|Q)$, is also transmitted with message $r_p$. The following constraints ensure that Bob can decode $r_c$, $(r_p,r_f)$ from his channel output $Y$ with an arbitrarily small probability of error:
\begin{IEEEeqnarray*}{rCl}
R_c			&<& k I(Q;Y)	\ ,\\
R_p + R_f	&<& k I(X;Y|Q)	\ .
\end{IEEEeqnarray*}

% - - - - - - - - - - - - - - - - - - - - - - - - - - - - - - - - - - - - - - -
\subsubsection*{Distortion at Bob}
Provided the above constraints are verified, Bob can decode $V$ with an arbitrarily small probability of error, and compute an estimate $\hat A$ of $A$ with mean distortion $\bE[ d(A,\hat A(V,B)) ]$.

% - - - - - - - - - - - - - - - - - - - - - - - - - - - - - - - - - - - - - - -
\subsubsection*{Equivocation Rate at Eve}
After some algebraic manipulations, it can be proved that the proposed scheme achieves any equivocation rate verifying the following inequality:
\[
\Delta	\leq	H(A|UE) - R_2 + R_p + R_f - k I(X;Z|Q) \ .
\]
The proof (which is omitted here due to the lack of space) follows the arguments of both~\cite[Section~IV-A]{villard2010secure}, and~\cite[Section~2.3]{liang2009information}, and relies on relation $r_1 = M'(r_c)$.

% - - - - - - - - - - - - - - - - - - - - - - - - - - - - - - - - - - - - - - -
\subsubsection*{End of Proof}
Putting all inequalities together, using Fourier-Motzkin elimination,
and prefixing an arbitrary DMC $P(X|T)$ to the DMC $P(Y,Z|X)$
prove Theorem~\ref{th:inner_bound}.
\endproof

%==============================================================================
\section{Sketch of Proof of Theorem~\ref{th:outer_bound} (Outer Bound)}
\label{sec:outer_bound}

Due to the lack of space, we only provide some of the basic
ideas underlying the proof of Theorem~\ref{th:outer_bound}. 
Details will be provided in an extended version of this paper.

For each $i\in\{1,\dots,n\}$ (resp. each $j\in\{1,\dots,m\}$), define the source (resp. channel) auxiliary random variables $U_i$, $V_i$ (resp. $Q_j$, $T_j$) as 
\begin{IEEEeqnarray*}{rCl}
U_i &=& (\phantom{A^{i-1},B^{i-1}, } B_{i+1}^n,E^{i-1}, Z^m) 	\ ,\\
V_i &=& (A^{i-1},B^{i-1},			 B_{i+1}^n,E^{i-1}, Y^m)	\ ,\\[.15cm]
Q_j &=& (\phantom{A^n, }	B^n, Y^{j-1}, Z_{j+1}^m) 			\ ,\\
T_j &=& (A^n,				B^n, Y^{j-1}, Z_{j+1}^m) 			\ .
\end{IEEEeqnarray*}
Note that $(U_i,V_i)\mkv A_i\mkv (B_i,E_i)$, and $Q_j\mkv T_j\mkv X_j\mkv (Y_j,Z_j)$ form Markov chains.

% - - - - - - - - - - - - - - - - - - - - - - - - - - - - - - - - - - - - - - -
\subsubsection*{Rate}
Using the chain rule for conditional mutual information, the Markov chain $(A_i,Y^m) \mkv (A^{i-1},B^n) \mkv E^{i-1}$, and the fact that random variables $A_i$, $B_i$, and $E_i$ are independent across time, we can prove that $I(A^n;Y^m|B^n) = \sum_{i=1}^n I(A_i;V_i | B_i)$.

From the chain rule, and the non-negativity of mutual information, we can also prove the following upper bound:
$I(A^n;Y^m|B^n) \leq	\sum_{j=1}^m I(T_j ; Y_j)$.

The above equations yield 
\[
\sum_{i=1}^n I(A_i;V_i | B_i) \leq \sum_{j=1}^m I(T_j ; Y_j)\ .
\]

% - - - - - - - - - - - - - - - - - - - - - - - - - - - - - - - - - - - - - - -
\subsubsection*{Distortion at Bob}
Bob reconstructs $g(Y^m,B^n)$. 
The $i$-th coordinate of this estimate is
$g_i(Y^m,B^{i-1},B_i,B_{i+1}^n) \triangleq	\hat A_i(V_i,B_i)$.
The component-wise mean distortion at Bob thus writes:
\[
\bE\big[ d(A^n,g(Y^m,B^n)) \big]
	=	\frac1n \sum_{i=1}^n \bE\left[ d(A_i,\hat A_i(V_i,B_i)) \right] \ .
\]

% - - - - - - - - - - - - - - - - - - - - - - - - - - - - - - - - - - - - - - -
\subsubsection*{Equivocation Rate at Eve}
From the chain rule for conditional entropy, and the Markov chain $A_i \mkv (A_{i+1}^n,E^i,Z^m)\linebreak \mkv (B_{i+1}^n,E_{i+1}^n)$, we can prove the following upper bound on the equivocation at Eve:
\[
H(A^n | E^n Z^m) \leq \sum_{i=1}^n H(A_i | U_i E_i) \ .
\]

Using the Markov chain $B^n\mkv A^n\mkv Z^m$, we expand the equivocation at Eve as follows:
\[
H(A^n | E^n Z^m)
	=	\underbrace{ I(A^n ; Y^m | B^n) - I(A^n ; Z^m | B^n)					}_{\Delta_c}
	+	\underbrace{ H(A^n | B^n Y^m) + I(A^n ; B^n | Z^m) - I(A^n ; E^n | Z^m)	}_{\Delta_s} .
\]

Following~\cite[Section~V]{csiszar1978broadcast},~\cite[Section~2.4]{liang2009information}, we can prove that $\Delta_c = \sum_{j=1}^m  I(T_j;Y_j|Q_j) - I(T_j;Z_j|Q_j)$, and following~\cite[Section~IV-B]{villard2010secure}, 
$\Delta_s = \sum_{i=1}^n H(A_i|V_i B_i) + I(A_i;B_i|U_i) - I(A_i;E_i|U_i)$.

% - - - - - - - - - - - - - - - - - - - - - - - - - - - - - - - - - - - - - - -
\subsubsection*{End of Proof}
Following the usual technique, we define independent random variables $K$, and $J$,
uniformly distributed over the sets $\{1,\dots,n\}$, and $\{1,\dots,m\}$, respectively. We also define random variables $A=A_K$, $B=B_K$, $E=E_K$, $U=(K,U_K)$, $V=(K,V_K)$, $X=X_J$, $Y=Y_J$, $Z=Z_J$, $Q=(J,Q_j)$, and $T=(J,T_j)$.
$(U,V)\mkv A\mkv (B,E)$ and $Q\mkv T\mkv X\mkv (Y,Z)$ still form Markov chains. 
Using these definitions, we prove the three inequalities of Theorem~\ref{th:outer_bound}.
Since they only involve \emph{marginal} distributions of auxiliary  variables, w.r.t. corresponding source/channel variables \emph{i.e.}, $p(uv|a)$ and $p(qt|x)$, we can define new auxiliary variables $\tilde U$, $\tilde V$, $\tilde Q$, and $\tilde T$, with identical marginal distributions, such that the (global) joint distribution writes $p(uvqtabexyz) = p(uv|a)p(abe)p(q|t)p(t|x)p(xyz)$ \emph{i.e.}, source and channel variables are independent.
\endproof

%==============================================================================
\section{Application Example and Discussion}
\label{sec:example}

%::::::::::::::::::::::::::::::::::::::
\begin{figure}
\centering
\begin{tikzpicture}[scale=.9]
	\node	(A) 	at (0,.5) 	{$A$};
	\node	(A0) 	at (0,0) 	{\small $0$};
	\node	(A1)	at (0,-2)	{\small $1$};
	
	\node	(B) 	at (-3,.5) 	{$B$};
	\node	(B0) 	at (-3,0) 	{\small $0$};
	\node	(Be)	at (-3,-1)	{\small $e$};
	\node	(B1)	at (-3,-2)	{\small $1$};
	
	\node	(E) 	at (3,.5) 	{$E$};
	\node	(E0) 	at (3,0) 	{\small $0$};
	\node	(E1)	at (3,-2)	{\small $1$};
	
	\draw[->] (A0) to node[midway,above]		{\small $1-\epsilon$}	(E0);
	\draw[->] (A0) to node[near start,above]	{\small $\epsilon$}		(E1);
	\draw[->] (A1) to node[near start,below]	{\small $\epsilon$}		(E0);
	\draw[->] (A1) to node[midway,below]		{\small $1-\epsilon$}	(E1);
	\draw[->] (A0) to node[midway,above]		{\small $1-\beta$}		(B0);
	\draw[->] (A0) to node[near start,below]	{\small $\beta$}		(Be);
	\draw[->] (A1) to node[near start,above]	{\small $\beta$}		(Be);
	\draw[->] (A1) to node[midway,below]		{\small $1-\beta$}		(B1);
\end{tikzpicture}
\caption{Binary erasure/binary symmetric side informations.}
\label{fig:ex_SI}
\end{figure}
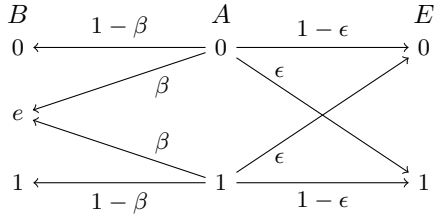
%::::::::::::::::::::::::::::::::::::::

Consider the source model depicted in Fig.~\ref{fig:ex_SI}, where the source is binary and the side information at Bob, resp. Eve, is the output of a binary erasure channel (BEC) with erasure probability $\beta\in[0,1]$, resp. a binary symmetric channel (BSC) with crossover probability $\epsilon\in[0,1/2]$, with input $A$.
The communication channel is similar to the one of~\cite{wyner1975wire}: It consists of a noiseless channel from Alice to Bob, and a BSC with crossover probability $\zeta\in[0,1/2]$, from Alice to Eve.

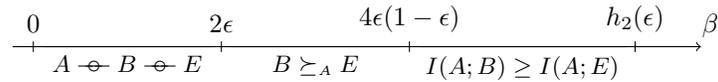
\begin{figure}[h]
\centering
\begin{tikzpicture}
	\node	(ci)	at (-.4,0)	{};
	\node	(c0) 	at (0,0) 	{};
	\node	(c1)	at (2.5,0)	{};
	\node	(c2)	at (5,0)	{};
	\node	(c3)	at (8,0)	{};
	\node	(e) 	at (9,0) 	{};
	
	\draw[->] (ci) to (e);
	\draw (c0) to node[anchor=north] {\small $A\mkv B\mkv E$}		(c1);
	\draw (c1) to node[anchor=north] {\small $B\lessnoisy{A}E$}		(c2);
	\draw (c2) to node[anchor=north] {\small $I(A;B)\geq I(A;E)$}	(c3);
	
	\draw (c0)+(0,-2pt) -- +(0,2pt)	node[anchor=south]	{$0$};
	\draw (c1)+(0,-2pt) -- +(0,2pt)	node[anchor=south]	{$2\epsilon$};
	\draw (c2)+(0,-2pt) -- +(0,2pt) 	node[anchor=south] 	{$4\epsilon(1-\epsilon)$};
	\draw (c3)+(0,-2pt) -- +(0,2pt) 	node[anchor=south] 	{$h_2(\epsilon)$};
	\draw (e) node[anchor=south] 	{$\beta$};
\end{tikzpicture}
\caption{Relative properties of the side informations as a function of $(\beta,\epsilon)$.}
\label{fig:cas}\vspace{-4mm}
\end{figure}

This model is of interest since neither Bob nor Eve can always be a lessnoisy decoder for all values of $(\beta,\epsilon)$.
Let $h_2$ denotes the binary entropy function given by $h_2(x) = -x\log_2(x) -(1-x)\log_2(1-x)$. According to the values of the parameters $(\beta,\epsilon)$, it can be shown by means of standard manipulations~\cite{nair2009capacitya} that the side informations satisfy the properties summarized in Fig.~\ref{fig:cas}.

%::::::::::::::::::::::::::::::::::::::
%::::::::::::::::::::::::::::::::::::::

From now on, let the distortion level at Bob be zero \emph{i.e.}, he performs \emph{lossless} reconstruction, and assume for simplicity that the source is uniformly distributed \emph{i.e.}, $\pr{A=0}=\pr{A=1}=1/2$. We focus on rate $k=1$ channel use per source symbol.
Under these assumptions, the inner bound of Theorem~\ref{th:inner_bound} is maximized by choosing $V=A$ and a uniformly distributed binary auxiliary random variable $U$ (resp. $Q$), produced as the output of a BSC with crossover probability $u\in[0,1/2]$ (resp. $q\in[0,1/2]$), and input $A$ (resp. $X$), as stated by the following proposition (which proof is omitted due to the lack of space).

\begin{prop}
\label{prop:example}
In the case considered in this section, region $\cR_\text{in}$ reduces to the set of all tuples $(k=1,D=0,\Delta)$ such that there exist $u,q\in[0,1/2]$ satisfying
\begin{IEEEeqnarray*}{rCl}
\beta(1-h_2(u))	&\leq& 1-h_2(q)	 								\ ,\\
\Delta	 		&\leq&  h_2(\epsilon)+h_2(u)-h_2(\epsilon\star u)	
						- \Big[ \beta h_2(u)				
						- \Big( h_2(\zeta) + h_2(q) - h_2(\zeta\star q) \Big) \Big]_+ ,
\end{IEEEeqnarray*}
where $a\star b = a(1-b) + (1-a)b$ for each $a,b\in[0,1]$.
\end{prop}

Notice that if $\beta\leq 4\epsilon(1-\epsilon)$ then $B\lessnoisy{A}E$, and hence Proposition~\ref{prop:B_less_noisy} holds \emph{i.e.}, the above inner bound is optimal. 

\subsection*{Counterexample for the optimality of Theorem~\ref{th:inner_bound}}

Let now assume that Bob does not have any side information \emph{i.e.}, $\beta=1$, and let $\epsilon=\zeta=0.1$ so that $A\mkv E\mkv B$ form a Markov chain, and neither Proposition~\ref{prop:B_less_noisy}, nor Proposition~\ref{prop:Z_less_noisy} applies. This setting provides a  counterexample for the general optimality of the inner bound in Theorem~\ref{th:inner_bound}. 
Numerical optimization over $u$ and $q$ in Proposition~\ref{prop:example} indicates that the proposed scheme achieves an equivocation rate $\Delta= 0.056$, while a naive analogue scheme consisting of directly plugging the source on the channel achieves $\Delta = 0.258$.
Furthermore, the latter concides with the outer bound of Theorem~\ref{th:outer_bound}.

The above example shows that a naive \emph{joint} source-channel scheme may achieve better performance in some cases. At first look, this is not surprising since it is well-known that \emph{joint} source-channel coding/decoding is a must for broadcast channels without secrecy constraints~\cite{gastpar2003code}, \cite{tuncel2006slepian-wolf}. However, the \emph{secure} setting is rather different because Alice only wants to help one receiver (Bob), while she wants to blur the other one (Eve). Therefore, the intuition indicates that  the optimal strategy  would be the opposite \emph{i.e.}, separation between source and channel encoders, as in Propositions~\ref{prop:B_less_noisy} and~\ref{prop:Z_less_noisy}.

%%%%%%%%%%%%%%%%%%%%%%%%%%%%%%%%%%%%%%%%%%%%%%%%%%%%%%%%%%%%%%%%%%%%%%%%%%%%%%%
%%%%%%%%%%%%%%%%%%%%%%%%%%%%%%%%%%%%%%%%%%%%%%%%%%%%%%%%%%%%%%%%%%%%%%%%%%%%%%%
\bibliographystyle{IEEEtran}
\bibliography{isit2011}

\end{document}